\documentclass[twocolumn,pra,superscriptaddress,noeprint]{revtex4-1}
\usepackage{graphicx}
\usepackage{amssymb,amsmath}
\usepackage{bm}
\usepackage{dcolumn}
\usepackage{subfigure}
\usepackage{float}
\usepackage{url}
\usepackage{xcolor}
\usepackage{ulem}
\usepackage{sidecap}
\usepackage{bbold}
\usepackage{booktabs}
\allowdisplaybreaks
\usepackage{hyperref}
\hypersetup{backref,pdfpagemode=FullScreen,colorlinks=true,breaklinks,urlcolor=blue,linkcolor=blue,citecolor=blue}

\usepackage{mathrsfs}

\begin{document}
\title{Manipulation of Spin Dynamics by Deep Reinforcement Learning Agent}

\author{Jun-Jie Chen}
\email{chen-jj13@mails.tsinghua.edu.cn}
\affiliation{State Key Laboratory of Low Dimensional Quantum Physics, Department of Physics, Tsinghua University, Beijing 100084, China}

\author{Ming Xue}
\affiliation{State Key Laboratory of Low Dimensional Quantum Physics, Department of Physics, Tsinghua University, Beijing 100084, China}





\begin{abstract}
We implement the reinforcement learning agent in spin-1 atomic system to prepare twin-Fock state from given initial state. Proximal policy gradient (PPO) algorithm is used to deal with continuous space of control field and the final optimized protocol is given by a stochastic policy. In both mean-field system and two-body quantum system, RL agent finds the optimal policies. In many-body quantum system, it also gives polices that outperform purely greedy policy and optimized adiabatic passage. These polices given by RL agent have good physical interpretability in phase space and may help us to understand the quantum dynamics. With thorough exploration of state space, RL policy is also robust to noises and have good generalization capability. In fact, RL could be highly versatile in quantum optimal control problems.
\end{abstract}

\maketitle

\section{introduction}
	Understanding the dynamics in quantum system is an important topic in physics. It is conductive to the optimal control of target state preparation in quantum information and precision measurement. The challenge is to find an optimal or sub-optimal protocol of external control field that can evolve the initial state to target state both quickly and accurately. Various methods based on quantum optimal control theory \cite{PhysRevA.92.062110,PhysRevA.85.042331,Brif_2010,beltrani2011exploring,PhysRevA.89.063408,PhysRevA.73.053401,PhysRevA.74.012721} or adiabatic shortcut \cite{PhysRevA.94.043623,PhysRevLett.114.177206,PhysRevA.93.023815,PhysRevA.97.012333,SelsE3909} are applied in different system to achieve better performance even reach the quantum speed limit. Generally, in a large class of linear quantum system, it is shown that the transition probability landscape has no local sub-optimal when the system is fully controllable \cite{Rabitz1998}, which means that perfect control can be find through traditional convex optimization such as gradient decent. However, the prerequisite of full controllability can be easily violated due to limitation of control field and discretion of time, $etc.$, and thus the original landscape crashes. It is also hard to implement such algorithm for large system due to computational complexity. On the other hand, all these methods are based on pure theories instead of experiences data, which sets a gap between simulation and true experiments.
	
	Reinforcement learning (RL) is a class of optimization algorithm that can learn an (sub)optimal policy from interaction with environment. By constantly observe the state of environment, take action and get feedback reward from it, RL agent collects the experiences data and use them to update its policy such that some long-term cumulative rewards are maximized. Compared to traditional optimal control theory, RL has two distinct advantages. First, RL can be implemented in a model-free way, i.e., the agent needs no prior human knowledge of the given system, which makes RL an universal learning framework for many dynamical systems. Second, RL can handle any given object function once we can design a proper reward in the problem and it provides great flexibility to achieve various optimization goals.
	
	In recent years, along with the development of deep learning, RL has achieved great progress in many areas including video and board games, natural language processing, electronic trading, $etc$ \cite{mnih2015human,silver2016mastering,silver2017mastering}. In physics, RL is gradually being widely used in quantum state preparation \cite{bukov2018reinforcement,yu2018reconstruction}, quantum computation and error correction \cite{andreasson2018quantum,herbert2018using,lin2018reinforcement,nautrup2018optimizing,sweke2018reinforcement}, quantum phase transitions \cite{bukov2017machine} and quantum robotics \cite{PhysRevX.4.031002}, $etc$. For various tasks, RL shows a certain degree of advantages over traditional theories and algorithms. Most works use value-based methods, such as Q-learning, and vanilla policy gradients to deal with dynamical systems with discrete action space, that is the value of control field is discretized. In fact, such discretization could change the landscape and makes the agent unable to learn true policy. The final results may also lack interpretability, even being nonphysical.
	
	In this work, we consider to deploy a RL agent on spin-1 atomic system which learns to generate twin-Fock state evolved from given initial state by controlling external magnetic field. This system is widely studied both theoretically and experimentally \cite{PhysRevA.47.5138,PhysRevA.66.033611,Lucke773,gross2011atomic,PhysRevLett.107.210406,Luo620}. In these works, spin squeezing is usually generated by using collective Rabi oscillation or adiabatic passage and high squeezing ratio can be realized. Here we use proximal policy optimization algorithm (PPO) to learn a better control protocol by numerical simulation. PPO belongs to actor-critic type RL algorithm proposed by OpenAI in \cite{schulman2017proximal} and has been applied in many challenging problems, such as complicated real-time strategy game and robotics. PPO can easily handle both continuous state and action space, which is also more realistic in physical systems. This work presents a general scheme converting a physical dynamical system to a standard RL task with proper state features, action representation and reward function. In section \ref{seciton-II}, we give a brief introduction to RL from scratch and specify the problems we consider. Section \ref{section-III} and \ref{section-IV} shows the learning policies of mean-field and quantum dynamical systems with different particle number. Section \ref{section-V} summarized the results and identify some problems for further improvement.
	
\section{Spin dynamics as a reinforcement learning task} \label{seciton-II}
\begin{figure}[!htbp]
\includegraphics[scale=0.45]{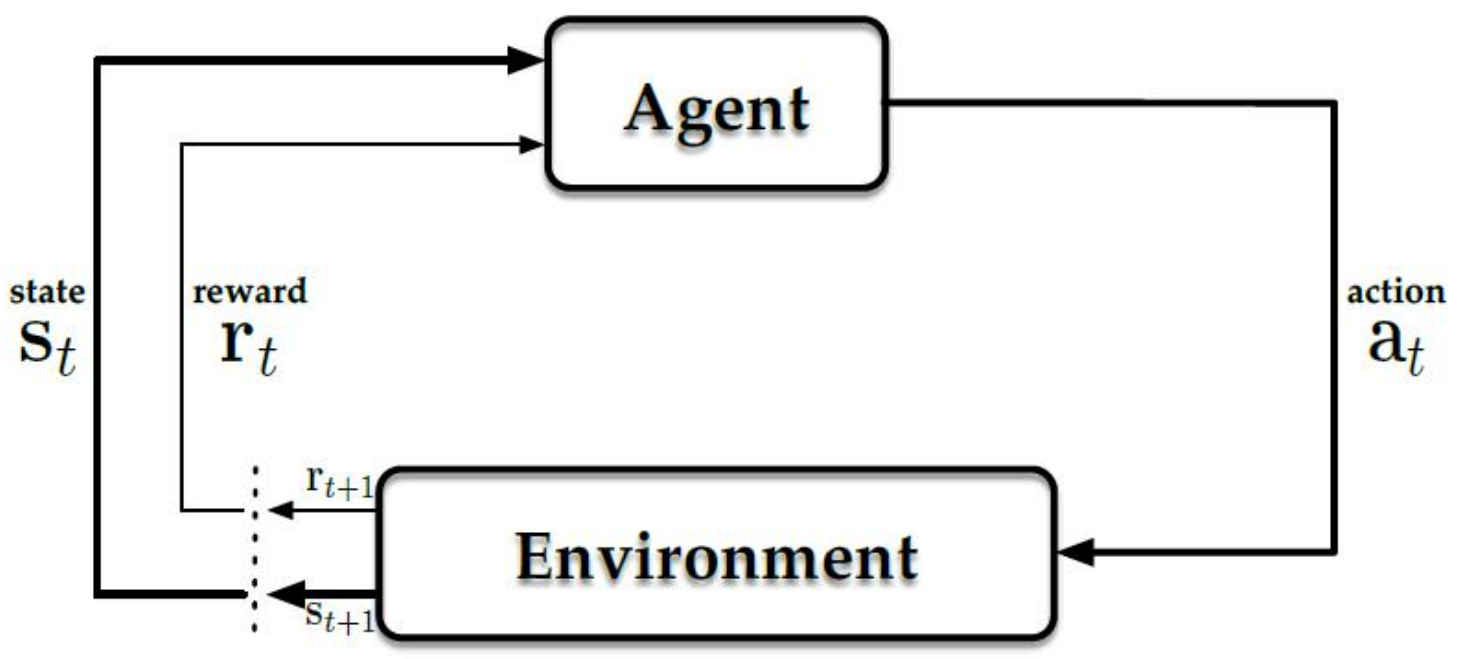}
\caption{Framework of typical MDP in reinforcement learning.}
\label{RLframe}
\end{figure}

	Reinforcement learning (RL) is an area of machine learning that learns how to act in a large class of dynamical system to maximize given cumulative rewards. Typical RL task is always modeled as a Markov decision process (MDP) as shown in Figs.(\ref{RLframe}). Such MDP is composed of two major objects, agent and environment. At each time $t$, the agent observes the environment and obtains its state feature $s_t$. Based on the current policy $\pi$ and state $s_t$, the agent takes action $a_t$ and acts on the environment. Then the environment will evolve to state $s_{t+1}$ due to $a_t$ and a reward $r_t$ will be fed back to agent itself. According to $r_t$ and previous experiences, the agent will update its policy $\pi$ by specific RL algorithm. The ultimate object for the agent is to learn an optimal policy $\pi^*$ that satisfies
\begin{eqnarray}
\pi^* = \mathop{\arg\max}_{\pi}J~~\text{with}~~J=\sum_{t=0}^{T_c}\gamma ^t\cdot r_t. \label{pi*}
\end{eqnarray}
Object function $J$ is a cumulative rewards with discounted factor $\gamma\in[0,1]$. When $\gamma=0$, the agent is totally greedy that only tries to maximize instantaneous reward $r_t$ at each time step. Typical $\gamma$ is always chosen to be closed to $1$ such that agent can find the true global optimal solution.

	In RL framework, the policy $\pi$ is a function 
\begin{eqnarray}
\pi:\mathcal{S}\rightarrow \mathcal{A},
\end{eqnarray}	
that maps the state space onto action space. Common policies are of two types, deterministic and stochastic. The deterministic policy is a function of state as $a=\pi(s)$, i.e., for given state $s$, the policy will give an unique value of action $a$. While the stochastic policy can be represented as $\pi(a|s)$, i.e., for given state $s$ the policy returns a distribution on action space $\mathcal{A}$. In fact, deterministic policy can be viewed as a special stochastic policy with zero variance around $a$. In deep RL, we usually use neural network or other parametric model to approximate the policy function as $\pi_{\theta}(s)$ and $\pi_{\theta}(a|s)$ with $\theta$ being trainable parameters. Then the RL task described in (\ref{pi*}) can be viewed as a general quadratic optimization problem and $\theta$ can be found by using gradient descent
\begin{eqnarray}
\theta \leftarrow \theta - \alpha \cdot \nabla_{\theta} J(\theta),
\end{eqnarray}
where $\alpha$ is learning rate. The gradient of $J$ can be further represented in a maximum likelyhood form as
\begin{eqnarray}
\nabla_{\theta}J(\theta)=\mathop E\limits_{\tau\sim\pi_{\theta}} \left[ \sum_{t=0}^{Tc}\nabla_{\theta}\log{\pi_{\theta}(a_t|s_t) A^{\pi_{\theta}}(s_t, a_t)} \right],
\end{eqnarray}
where $\tau$ is a trajectory following $\pi_{\theta}$ and $A^{\pi_{\theta}}$ is the advantage function. There are two ways to estimate the advantage function, statistic and parametric ways. Vanilla policy gradient algorithm uses statistic inference from experiences data $(s_t,a_t,r_t,s_{t+1})$ to estimate $A^{\pi_{\theta}}$. Actor-critic algorithm uses another neural network (or other parametric model) to approximate $A^{\pi_{\theta}}$. Here $\pi_{\theta}$ is the actor which is responsible for choosing action while $A^{\pi_{\theta}}$ is the critic which gives the value of state and action. In this work, we use proximal policy optimization algorithm (PPO) to learn a stochastic Guassian policy $\pi_{\theta}(a|s)$. PPO is an advanced actor-critic type algorithm developed in recent years. Compared to traditional methods, PPO has a more robust learning process due to first-order trust region search gradient descent and can handle both discrete and continuous action space. More details can be found in \cite{schulman2017proximal}.
	
	In this paper, we consider a spin-1 system with its Hamiltonian being
\begin{eqnarray}
H &=& \int d\vec{r}~\sum_{m=-1}^1 \hat{\psi}_m^{\dagger} \left[ -\frac{\hbar^2}{2m}\nabla^2+V-pm+q(t)m^2 \right]\hat{\psi}_m \nonumber \\
&& + \int d\vec{r}~\frac{1}{2}\left[ c_0:\hat{n}^2: + c_2:\hat{\bold{F}}^2: \right], \label{spin1-hamiltonian}
\end{eqnarray}
where $p$ is linear Zeeman and $q(t)$ is time-dependent quadratic Zeeman, $\hat{n}=\sum_m \hat{n}_m$ ($m=\pm1,0$) is total density operator and $\hat{F}_i=\sum_{m,n}(\hat{F}_i)_{mn}\hat{\psi}_m^{\dagger}\hat{\psi_n}$ ($i=x,y,z$) is the spin-1 operator. The Hamiltonian conserves total magnetic moment $F_z$ and particle number. In the following, we always consider the dynamics in $F_z=0$ subspace and ferromagnetic interaction with $c_2<0$. The ultimate goal is to find (a)an (sub-)optimal protocol $q(t)$ that evolves the quantum state from given initial state $|\psi_i\rangle$ to given target state $|\psi_f\rangle$. Here we choose the target state to be the twin-Fock state $|\psi_f\rangle=|N/2,0,N/2\rangle$ which is a squeezed state in spin space and very valuable for quantum precise measurements. The initial state $|\psi_i\rangle$ can be arbitrarily given.

Now we are ready to convert our problem of quantum state preparation in spin-1 system to a standard RL task. To this end, we will first specify the definition of $\mathcal{S},\mathcal{A},\mathcal{R}$ for our system.
\begin{itemize}
\item State space $\mathcal{S}$: The most straight-forward representation of $s_t$ for a quantum state is its wave function $|\psi\rangle$ because it contains all the information we need. However, $|\psi\rangle$ is not always the best choice, especially for a many-body quantum system since its dimension increase (exponentially) as particle number increasing. An alternative is to use representative physical observables to describe state $s_t$. Compared to wave function, there are two advantages. First, the dimension of state features is unchanged which makes it possible to generalize the policy to different particle number. Second, the output policy has more interpretability when state feature has clear physical meaning. In this work, we use the second option.
\item Action space $\mathcal{A}$: Action is clear in this problem that $a_t=q(t)$, the second-order quadratic Zeeman term at time $t$. Experimentally, $q$ can be tuned by external magnetic field or by microwave dressing.
\item Reward space $\mathcal{R}$: Reward $r_t$ selection depends on the optimization target itself. Here we want to achieve maximum fidelity at final time $T_c$, that is
\begin{eqnarray} 
J = |\langle \psi(T_c)|\psi_f\rangle|^2. \label{object_func}
\end{eqnarray}
However, feedback might be too sparse if we only give reward at final time $T_c$ and the training process will be very hard, even fail under current setup. In fact, we can decompose object function (\ref{object_func}) into a summation as
\begin{eqnarray}
J &=& \sum_{i=0}^n \left(|\langle \psi(t_i)|\psi_f\rangle|^2-|\langle \psi(t_{i-1})|\psi_f\rangle|^2\right).
\end{eqnarray}
Here, total evolution time $T_c$ is discretized into consecutive period ended at $t_i$ ($i=0,...n$). At each time step $t_i$, a reward
\begin{eqnarray}
r_i=|\langle \psi(t_i)|\psi_f\rangle|^2-|\langle \psi(t_{i-1})|\psi_f\rangle|^2,
\end{eqnarray}
which is the instantaneous change of fidelity, is fed back to agent. Dense rewards scheme makes the training more quickly and stable. Though $r_i$ seems to be greedy on fidelity increasing, RL algorithm always try to maximize $J$, i.e., the summation of $r_t$. It won't restrict us on greedy policy and RL agent still learns to find global optimal. We can further modified the reward as
\begin{eqnarray}
r_i = -\ln \left(\frac{1-|\langle \psi(t_i)|\psi_f\rangle|^2}{1-|\langle \psi(t_{i-1})|\psi_f\rangle|^2} \right),
\end{eqnarray}
which ensures the agent can still learn well even when the state has evolved to near target.
\end{itemize}

	In the following sections, we show the performance of RL agent on various environments, including mean-field and quantum dynamics. Some interpretabilities are also extracted from the final policies.

\section{Mean-field Dynamics}\label{section-III}

	Under the mean-field approximation, we have $\hat{\psi}=(\xi_1, \xi_0, \xi_{-1})^{T}$ with field parameterization $\xi_1=\sqrt{(1-\rho_0)/2}\cdot e^{i\chi_{+}}, \xi_0=\sqrt{\rho_0}$ and $\xi_{-1}=\sqrt{(1-\rho_0)/2}\cdot e^{i\chi_{-}}$. Here $\rho_0$ is the population density on $m_F=0$ state and $\chi_{\pm}$ is the phase of $m_F=\pm1$ components. Then the spin dynamics in $F_z=0$ subspace is governed by

\begin{eqnarray}
\dot{\rho_0} &=& \frac{2c_2}{\hbar}\rho_0(1-\rho_0)\sin{\theta_s}, \label{mf_eqn1}\\
\dot{\theta_s} &=& -\frac{2q}{\hbar}+\frac{2c_2}{\hbar}(1-2\rho_0) \left[ 1+\cos{\theta_s} \right], \label{mf_eqn2}
\end{eqnarray}
where $\theta_s=\chi_+ + \chi_-$. The solution of the coupled equations is a classical non-rigid pendulum, which is a typical RL benchmark task. Our target for this task is to evolve initial state to $\rho_0=0$ state. In this system, the state is $s_t=(\rho_0(t), \theta_s(t))$ and the action is $q$ itself. Since the phase space of mean-field dynamics is 2-dimensional, state tuple $(\rho_0, \theta_s)$ is sufficient to represent the whole system. That is, this system is a pure MDP.

\begin{table}
\caption{Mean-field system}
\begin{tabular}{cc}
\hline
Hyperparameters& value\\
\hline
hidden size& [32, 16] \\
activation& $\tanh$ \\
discounted factor $\gamma$& 0.999 \\
actor-network learning rate& 3E-4 \\
critic-network learning rate& 1E-3 \\
steps/episode& 100 \\
time/step($dt$)& 0.05 \\
target KL-divergence& 0.01 \\
\hline
\end{tabular} \label{HP-table-1}
\end{table}

Now we deploy PPO agent on this mean-field system. We choose a very small neural network to approximate the actor and critic. The hidden layer contains two fully-connected layers (or MLP) with 32 and 16 neurons on layer and the activation function is tanh. The output layer activation function is also $\tanh$ so that we can restrict $q$ being in the range $(q_{\min}, q_{\max})$. In this simulation, we choose $c_2=-1.0,\hbar=1$ and restrict $q$ to be in $(-6|c_2|,6|c_2|)$. The hyper-parameters are listed in the Table \ref{HP-table-1}. At least for such simple system, no sophisticated hyper-parameter tuning or post-selection is needed for our experiments.

\begin{figure}[!htbp]
\includegraphics[scale=0.36]{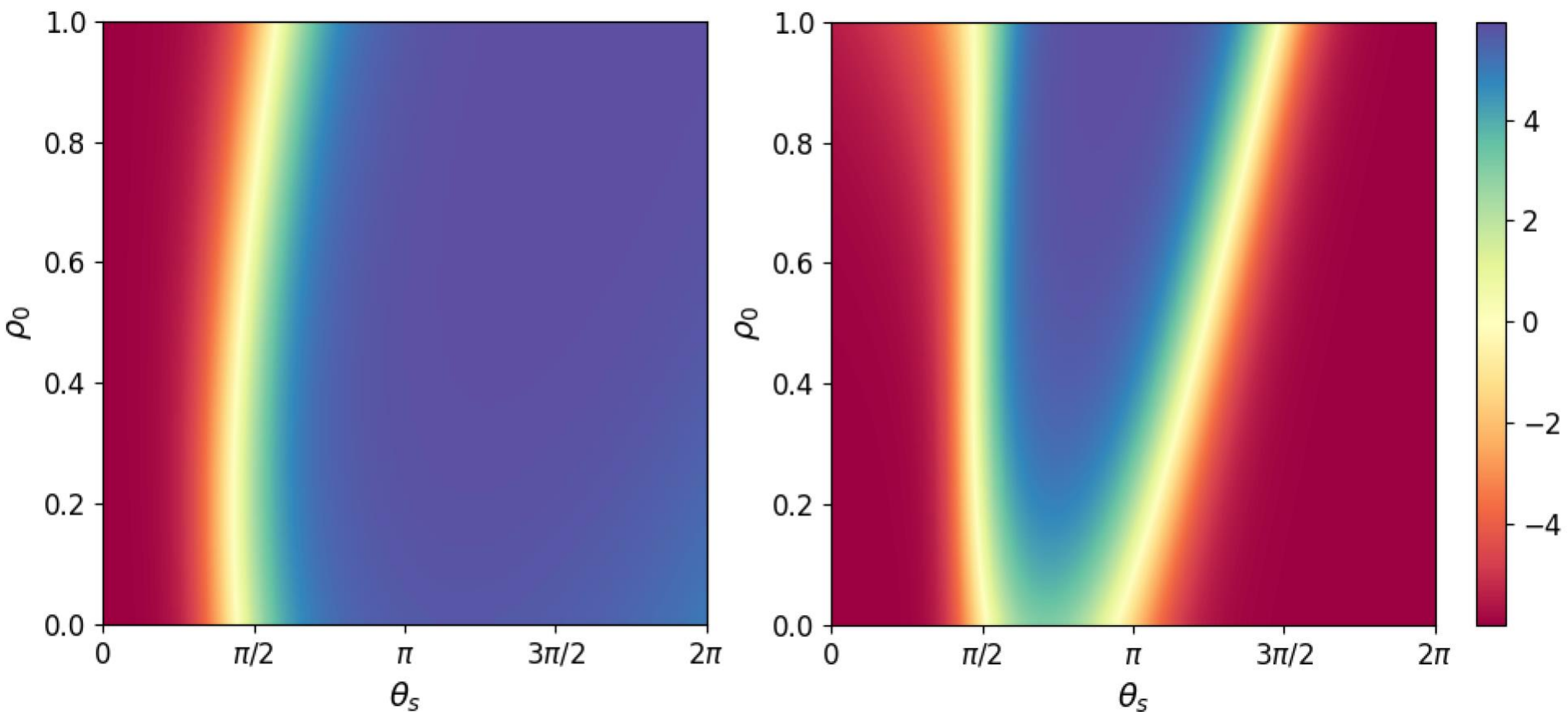}
\caption{Mean value of Guassian stochastic policy on phase space $(\theta_s, \rho_0)$. (left) $\pi_s$ use fixed initial state $(\theta_s, \rho_0)=(0.0, 0.9)$ for each training episode. (right) $\pi_g$ use random initial state for each training episode. Total training epochs number is 200.} \label{mf_policy}
\end{figure}

\begin{figure}[!htbp]
\includegraphics[scale=0.6]{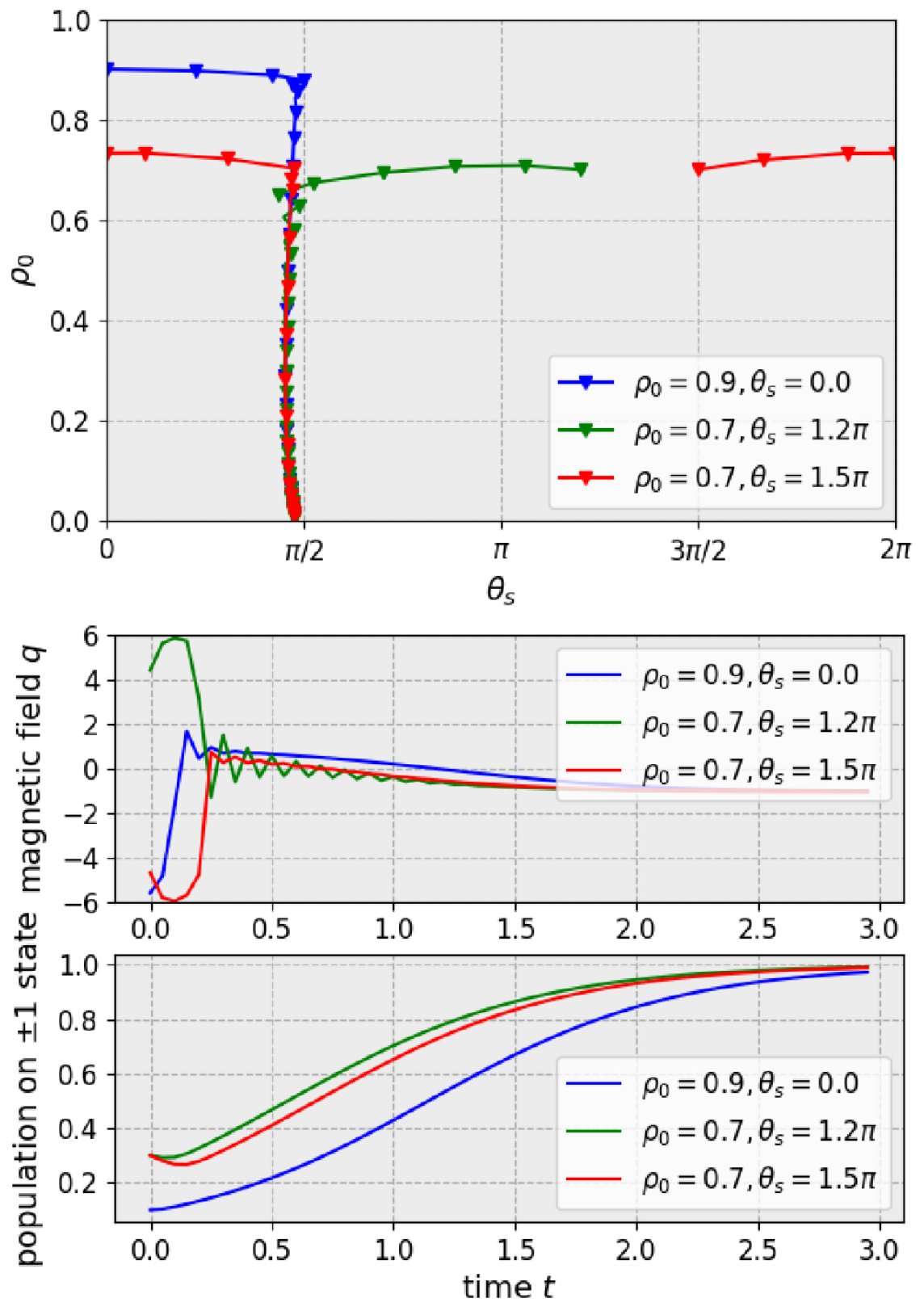}
\caption{Results of three typical initial states. (a) evolution trajectories on phase space guided by generalized policy $\pi_g$. (b) generalized policy $\pi_g$ of magnetic field $q(t)$ given by $\pi_g$. (c) time evolution of spin population on $m_F=\pm1$ state.} \label{mf_trajectory}
\end{figure}

In Figs.(\ref{mf_policy}), we show the two polices learned by the RL agent on phase space $(\theta_s, \rho_0)$. The first policy $\pi_s$ corresponds to fixed initial state setup. In each training episode, the initial state is always chosen to be $(\theta_s, \rho_0)=(0.0,0.9)$. The second policy $\pi_g$ corresponds to randomized setup, i.e. in each training episode, the initial state is chosen randomly among the whole phase space. These polices work as a map in phase space that identify the best $q$ value to take starting from arbitrary state $(\theta_s, \rho_0)$. Compared to each other, we observe that $\pi_g$ explore the phase space more thoroughly than $\pi_s$ due to random initialization. While $\pi_s$ only explore half of the phase space because it always start from fixed state and soon all sampled trajectories fall into the optimal region around $\theta_s\sim \pi/2$.

Figs.(\ref{mf_trajectory}) shows the evolution results of three typical initial states $(\theta_s,\rho_0)=(0.0,0.9),(1.2\pi,0.7)$ and $(1.5\pi,0.7)$ guided by policy $\pi_g$. In Figs.(\ref{mf_trajectory}a), we can see that all trajectories have the same characteristics. They always move horizontally first until they reach $\theta_s=\pi/2$ and then move along this geodesic path to the target $\rho_0=0$. Figs.(\ref{mf_trajectory}b) plots the time series of $q(t)$. At the beginning of evolution, $q$ is much larger than $c_2$ such that the spin dynamics is almost frozen while magnetic phase $\theta_s$ is accumulating rapidly, which corresponds to the horizontal trajectory in Figs.(\ref{mf_trajectory}a). Then $q$ decays to small value comparable with $c_2$ that activates spin dynamics and the population on $m_F=\pm1$ states increases. In fact, it is obvious to show that such policy is optimal. From RHS of (\ref{mf_eqn1}), the decay rate of $\rho_0$ is maximized only when $\theta=\pi/2$. To keep $\theta$ constant, we have to set RHS of (\ref{mf_eqn2}) to be zero all the time. We have
\begin{eqnarray}
q(t)= c_2(1-\rho_0(t)), \label{mf-optimal-sol}
\end{eqnarray}
which is quite similar to the protocol shown in Figs.(\ref{mf_trajectory}b) when $t>0.5$. In fact, if $q$ is unlimited, we can set $q=\pm\infty$ at the initial time and tune the magnetic phase $\theta_s$ to $\pi/2$ with no time cost. To sum up, in this simple mean-field spin dynamics system, the RL agent is able to learn an optimal policy that can evolve any initial state to target state. 

\section{Quantum Dynamics}\label{section-IV}

	The quantum dynamics of spin-1 system can be revealed from single mode approximation (SMA). Under SMA, we assume that all particles share the same spatial mode. This allows the field operator to be approximated as $\hat{\psi}_m=\hat{a}_m\cdot \phi(\vec{r})$ ($m=0,\pm1$) where $\phi$ is the wave function of spatial mode and $\hat{a}_m$ is the annihilation operator of spin state $m$. The hamiltonian under SMA is
\begin{eqnarray}
H &=& \frac{c_2}{2N}\left[ (2\hat{N}_0-1)(\hat{N}_1+\hat{N}_{-1}) \right. \nonumber \\
&& + \left. 2(\hat{a}_1^{\dagger}\hat{a}_{-1}^{\dagger}\hat{a}_0\hat{a}_0 + h.c.) \right]-q\hat{N}_0, \label{sma_hamiltonian}
\end{eqnarray}
where $N$ is total particle number and $c_2$ is the coupling strength of spin-exchange interaction. Still we assume $c_2=-1.0$ and total magnetic moment $F_z=0$ in the following discussion.

	In quantum system, we can directly use wave function as the representation of state $s_t$ which makes the spin dynamic a pure MDP. For particle number $N$, the Hilbert space has dimension $N/2+1$ and $s_t$ feature size is $N+2$ (module and phase). However, using wave function as features will lose some interpretability and generalization ability as discussed before. Alternatively, we use physical observables $\rho_0$ and $\theta_s$ as state features, in which
\begin{eqnarray}
\rho_0 &=& \langle \hat{N}_0 \rangle/N, \\
\theta_s &=& \text{args}~\langle \hat{a}_1^{\dagger}\hat{a}_{-1}^{\dagger}\hat{a}_0\hat{a}_0 \rangle.
\end{eqnarray}
These two observables are almost identical to those used in mean-field dynamics. In fact, all representative observables in this spin-1 quantum system, including total angular momentum $\langle\hat{L_i}\rangle$, $\langle\Delta\hat{L_i}\rangle$ ($i=x,y,z$) even squeezing ratio $\xi$, can be inferred from $\rho_0$ and $\theta_s$. On the other hand, $\rho_0$ and $\theta_s$ loss some information of the true quantum state which makes it a partially observable MDP (POMDP) and may have negative impacts on policy performance. The hyper-parameters used for our simulations are listed in Table-\ref{HP-table-2} and no sophisticated tuning is made either. The framework is quite similar to that used in mean-field system.

\begin{table}
\caption{Quantum system}
\begin{tabular}{cc}
\hline
Hyperparameters& value\\
\hline
hidden size& [64, 32] \\
activation& $\tanh$ \\
discounted factor $\gamma$& 0.999 \\
actor-network learning rate& 3E-4 \\
critic-network learning rate& 1E-3 \\
steps/episode& 200 \\
time/step($dt$)& 0.1 \\
target KL-divergence& 0.01 \\
\hline
\end{tabular} \label{HP-table-2}
\end{table}

\subsection{Two-body problem $\bold{N=2}$}
\begin{figure}[!htbp]
\includegraphics[scale=0.33]{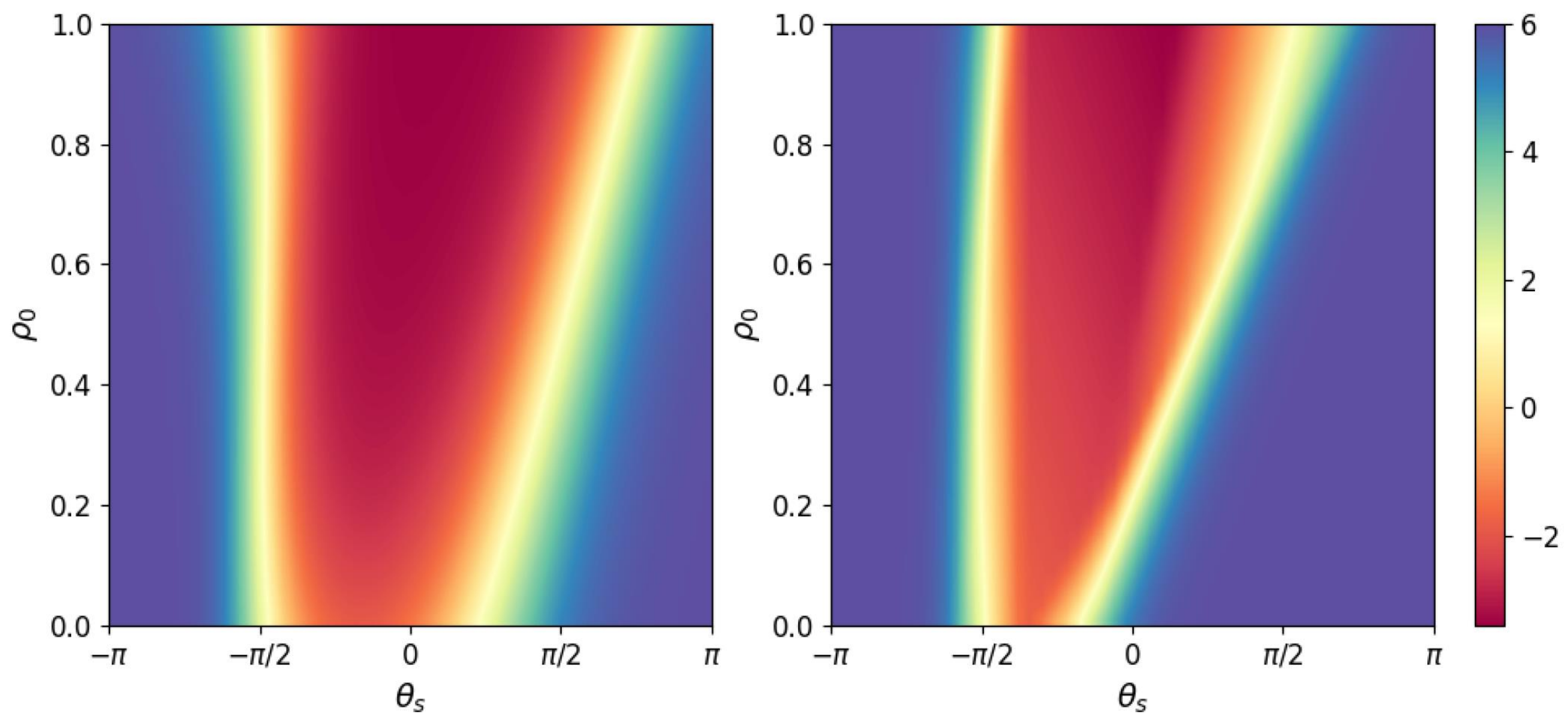}
\caption{Mean value of stochastic policy on phase space $(\theta_s, \rho_0)$. (left) $\pi_s$ use fixed initial state $\psi=|0,2,0\rangle$ for each training episode. (right) $\pi_g$ use random initial state for each training episode. Total training epochs number is 200.}
\label{sma_N2_policy}
\end{figure}

\begin{figure}[!htbp]
\includegraphics[scale=0.65]{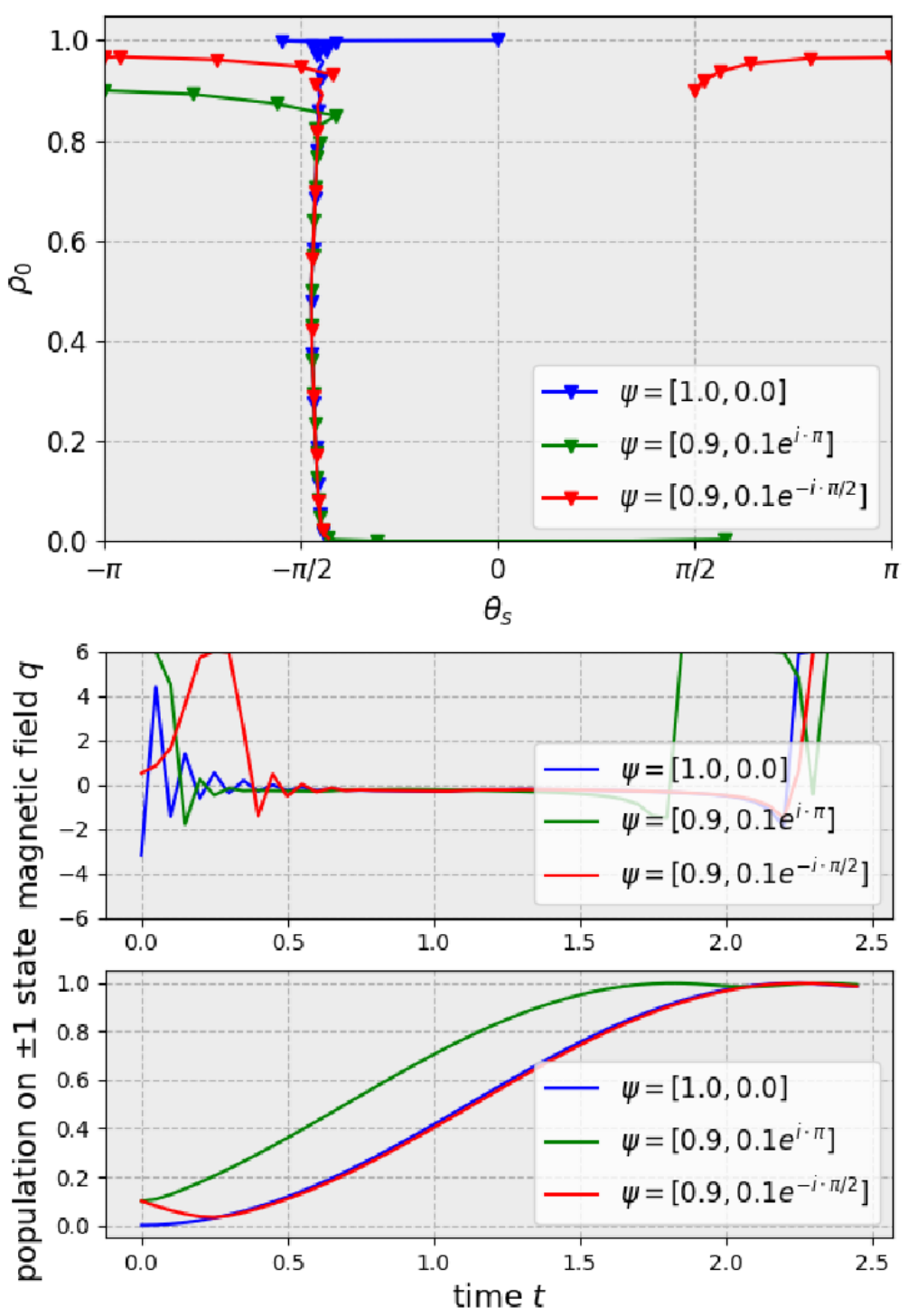}
\caption{Results of three typical initial states. (a) evolution trajectories on phase space guided by generalized policy $\pi_g$. (b) generalized policy $\pi_g$ of magnetic field $q(t)$ given by $\pi_g$. (c) time evolution of spin population on $m_F=\pm1$ state.}
\label{sma_N2_trajectory}
\end{figure}

First, we consider the simplest case with $N=2$. The wave function can be represented as a pseudo-spin-1/2 on Bloch sphere
\begin{eqnarray}
|\psi\rangle = \cos{\frac{\theta}{2}}|1\rangle + \sin{\frac{\theta}{2}}e^{i\phi}|2\rangle,
\end{eqnarray}
where $|1\rangle=|0,2,0\rangle, |2\rangle=|1,0,1\rangle$ and it is easy to show that $\rho_0=\cos^2 \theta/2$ and $\theta_s=-\phi$. The equation of motion is
\begin{eqnarray}
\dot{\theta} &=& -\sqrt{2}c_2\sin{\phi}, \label{QN2-eq1} \\
\dot{\phi} &=& -\frac{c_2}{2}+2q+\frac{\sqrt{2}}{2}\cos{\phi}\left( \tan{\frac{\theta}{2}}-\cot{\frac{\theta}{2}} \right). \label{QN2-eq2}
\end{eqnarray}
The optimal policy that evolves state $|1\rangle$ to $|2\rangle$ is obtained when $\phi\equiv\pi/2$ ($\theta_s=-\pi/2$) and $\dot{\phi}\equiv 0$, which implies
\begin{eqnarray}
q(t) = c_2/4. \label{N2-optimal-sol}
\end{eqnarray}
It is just a simple Rabi-oscillation between initial state $|0,2,0\rangle$ and target state $|1,0,1\rangle$. In Figs.(\ref{sma_N2_policy}), we show the policies learned by RL agent with fixed ($\pi_s$) or random ($\pi_g$) initial state. Here we also observe that the phase space (Hilbert space) is explored more thoroughly with randomized initial state. Figs.(\ref{sma_N2_trajectory}) shows the evolution results of three typical initial states $|\psi\rangle=|1\rangle,0.9|1\rangle+0.1e^{i\pi}|2\rangle$ and $0.9|1\rangle+0.1e^{-i\pi/2}|2\rangle$ based on policy $\pi_g$. The trajectories learned by the agent have same characteristics as those of mean-field system and magnetic field $q$ is closed to theoretical optimal value $c_2/4$ when $\theta_s$ is adjusted to near $-\pi/2$.

	We can further obtain the quantum speed limit $T_{\text{QSL}}$ of this simple system
\begin{eqnarray}
T_{\text{QSL}} = \min \int_0^{\pi} \frac{1}{\dot{\theta}}d\theta = \frac{\pi}{\sqrt{2}|c_2|}, \label{QSL}
\end{eqnarray}
which is almost achieved by our policy shown in Figs.(\ref{sma_N2_trajectory}-c). We also notice that equation($\ref{QSL}$) is identical to the Bhattacharyya bound $T_{\text{QSL}} = \Delta E_0^{-1}\arccos{\langle \psi_i|\psi_f\rangle}$ where $|\psi_i\rangle=|1\rangle$, $|\psi_f\rangle=|2\rangle$ and $\Delta E_0$ is the energy variance of initial state $|\psi_i\rangle$. In our case, $\Delta E_0=\sqrt{2}|c_2|/2$. This is obvious because the dimension of Hilbert space is only two when $N=2$. To sum up, RL agent is able to learn the optimal policy in simple two-body system.

\subsection{Many-body problem $\bold{N=10}$}
	In the following we consider the many-body quantum dynamics with particle number $N=10$. Now the system has two major differences from two-body and mean-field dynamics. First, the system becomes a POMDP when we use $s_t=(\rho_0, \theta_s)$. But since $\rho_0$ and $\theta_s$ catch most of the important features of this system, we may still learn a good policy. Second, when the dimension of Hilbert space gets larger, there will be a longer frozen time period at the early stage of evolution in which the fidelity on target state almost remains zero. This frozen time  It makes the reward being naturally sparse and training process could be hard. However, we still get applicable policies under such setup at least for many body system with tens of particle.
	
\begin{figure}[!htbp]
\includegraphics[scale=0.31]{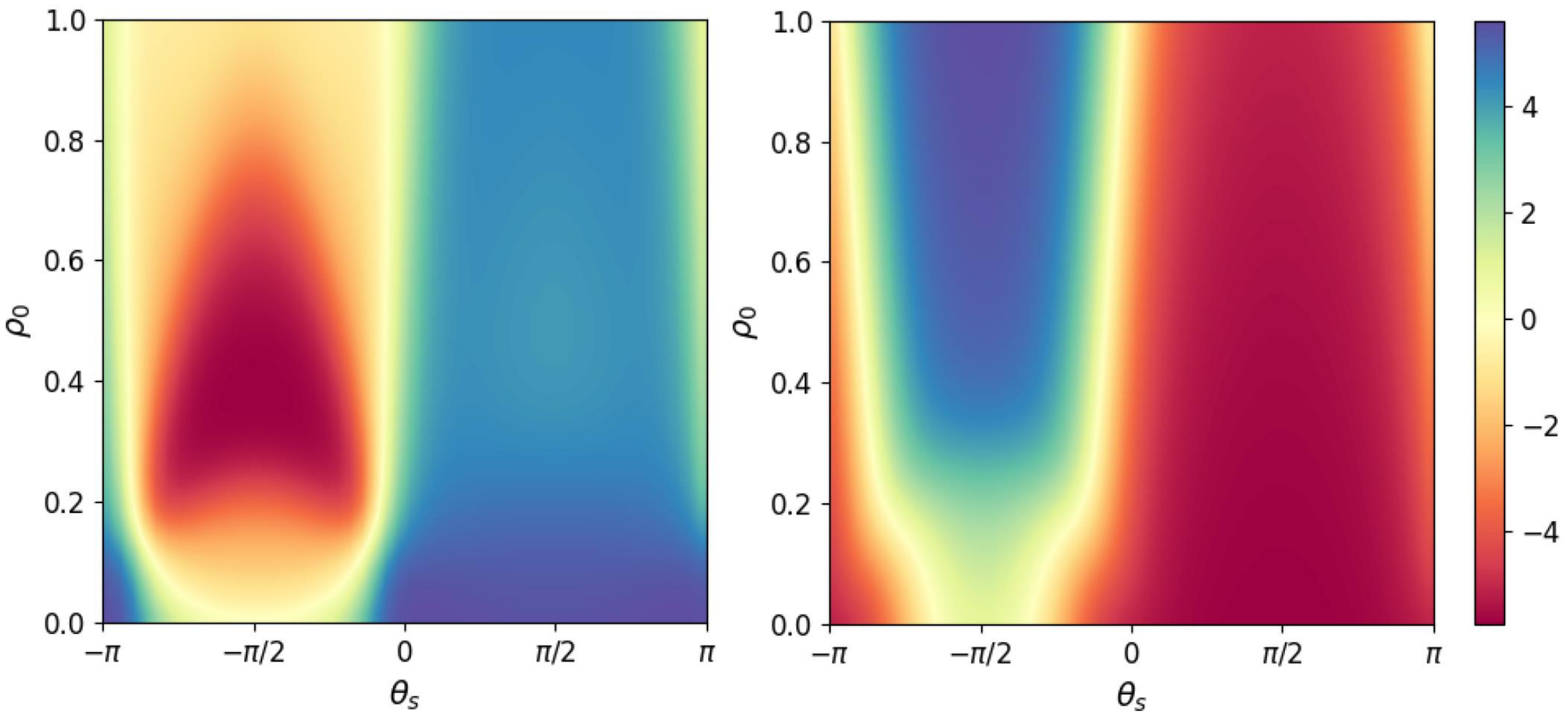}
\caption{Mean value of stochastic policy on phase space $(\theta_s, \rho_0)$. (left) $\pi_s$ use fixed initial state $\psi=|0,10,0\rangle$ for each training episode. (right) $\pi_g$ use random initial state for each training episode. Total training epochs number is 1000.}
\label{sma_N10_policy}
\end{figure}

\begin{figure*}[!htbp]
\centering
\includegraphics[scale=0.58]{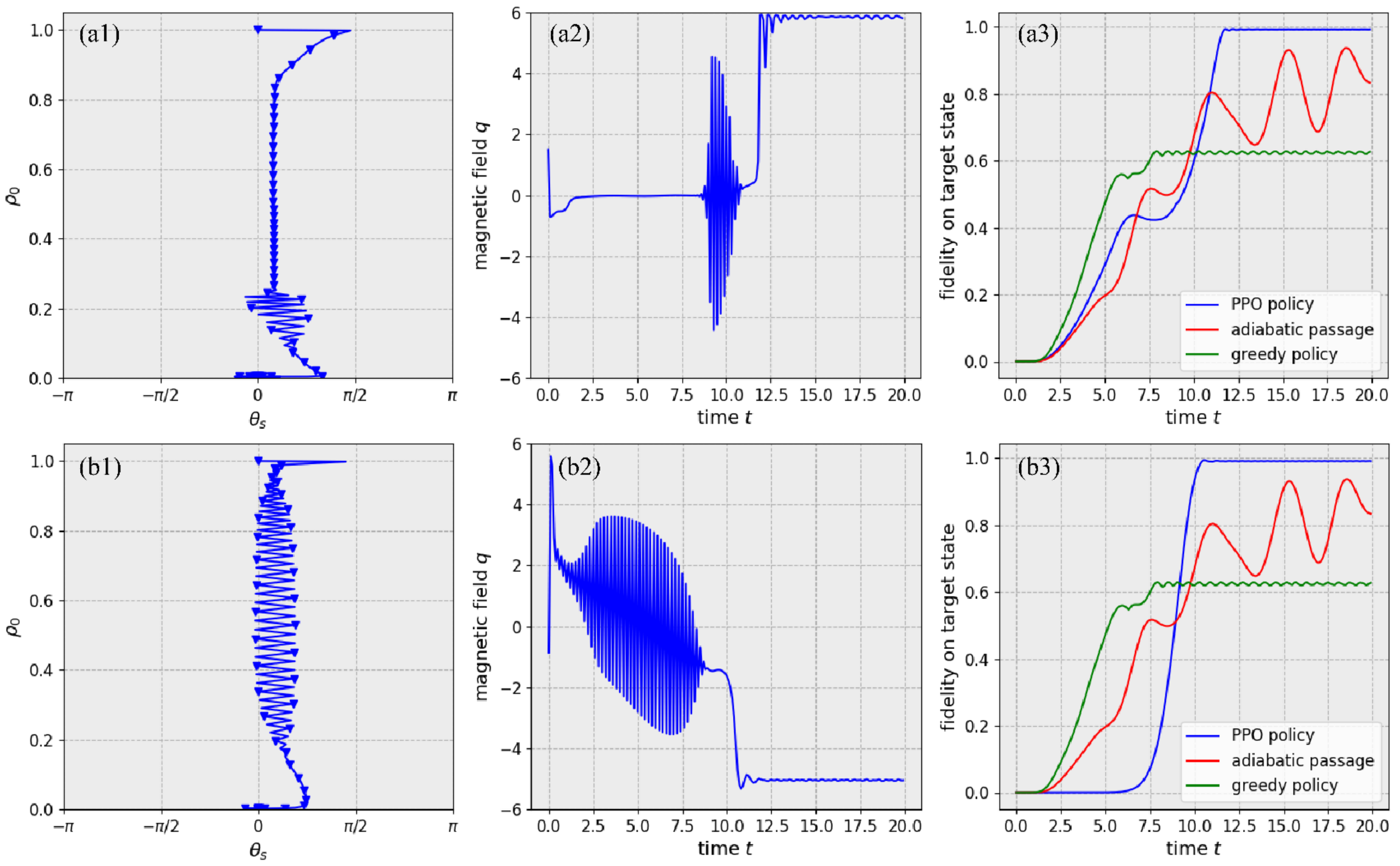}
\caption{Results of initial state $|0,N,0\rangle$ (a1,b1) evolution trajectory on phase space guided by policy $\pi_s$ or $\pi_g$. (a2, b2) magnetic field $q(t)$ given by $\pi_s$ or $\pi_g$. (a3) time evolution of fidelity on target state $|N/2,0,N/2\rangle$ following $\pi_s$ or $\pi_g$. Here the adiabatic passage refers to linear ramp in time period $[0,t]$ with $q$ changing from $q_i$ to $q_f$. $q_i$ and $q_f$ is optimized such that this linear ramp shows best performance. In greedy policy, $q$ is chosen at each time step to maximize instantaneous reward, that is the one-step fidelity increment.}
\label{sma_N10_trajectory}
\end{figure*}
	
	Figs.(\ref{sma_N10_policy}) shows the final polices learned by RL agent with fixed ($\pi_s$) and randomized ($\pi_g$) initial state. To ensure convergence, the number of training epochs now increases to 1000 due to the complexity of many-body system. Unlike mean-field or two-body system, $\pi_s$ and $\pi_g$ are quite different from each other. We further show evolution trajectories of quantum states that starts from $|0,N,0\rangle$ following $\pi_s$ and $\pi_g$ in Figs.(\ref{sma_N10_trajectory}) respectively. Policy $\pi_s$ combines Rabi-oscillation with magnetic phase angle precession. In early stage, when $t\lesssim 7.5$, control field $q\simeq -0.02$ which is a Rabi-oscillation process and at the end of this period, fidelity stops to increase. In the middle stage, when $t\lesssim 12.0$, field $q$ is oscillating such that magnetic phase $\theta_s$ can be tuned to proper value as soon as possible and fidelity can continuous to increase. In final stage, when $t\gtrsim 12.0$, the fidelity is almost unity ($>0.999$) and $q$ is tuned to max value to freeze the spin dynamics. Policy $\pi_g$ behaves similar to $\pi_s$ in phase space while the control protocol of $q$ is more aggressively. The RL agent of $\pi_g$ learns to tune $\theta_s$ at every time step and the final protocol is a trade-off between Rabi-oscillation and magnetic phase angle precession. In fact, $\pi_g$ can be treated as a optimization of adiabatic passage as $q$ decreases from $6.0$ to $-6.0$ in overall trend. We compare the performance of RL policies with optimized adiabatic linear ramp and purely greedy policy. Both $\pi_s$ and $\pi_g$ exceed these traditional methods. In fact, RL agent is definitely not greedy and always try to maximize long-term total rewards. This is corroborated by Figs.(\ref{sma_N10_trajectory}-a3,b3) in which PPO policy is slower (even frozen) than greedy one at early stage while faster later. The agent learns how to weigh short-term and long-term benefits. In comparison, $\pi_g$ is better than $\pi_s$ with shorter evolution time to unity fidelity. This is due to the fact that $\pi_g$ is learned with random initial state. Thus the Hilbert space is explored more thoroughly in $\pi_g$ which avoids letting the agent falls into bad local optimal.
	
\begin{figure*}[!htbp]
\centering
\includegraphics[scale=0.9]{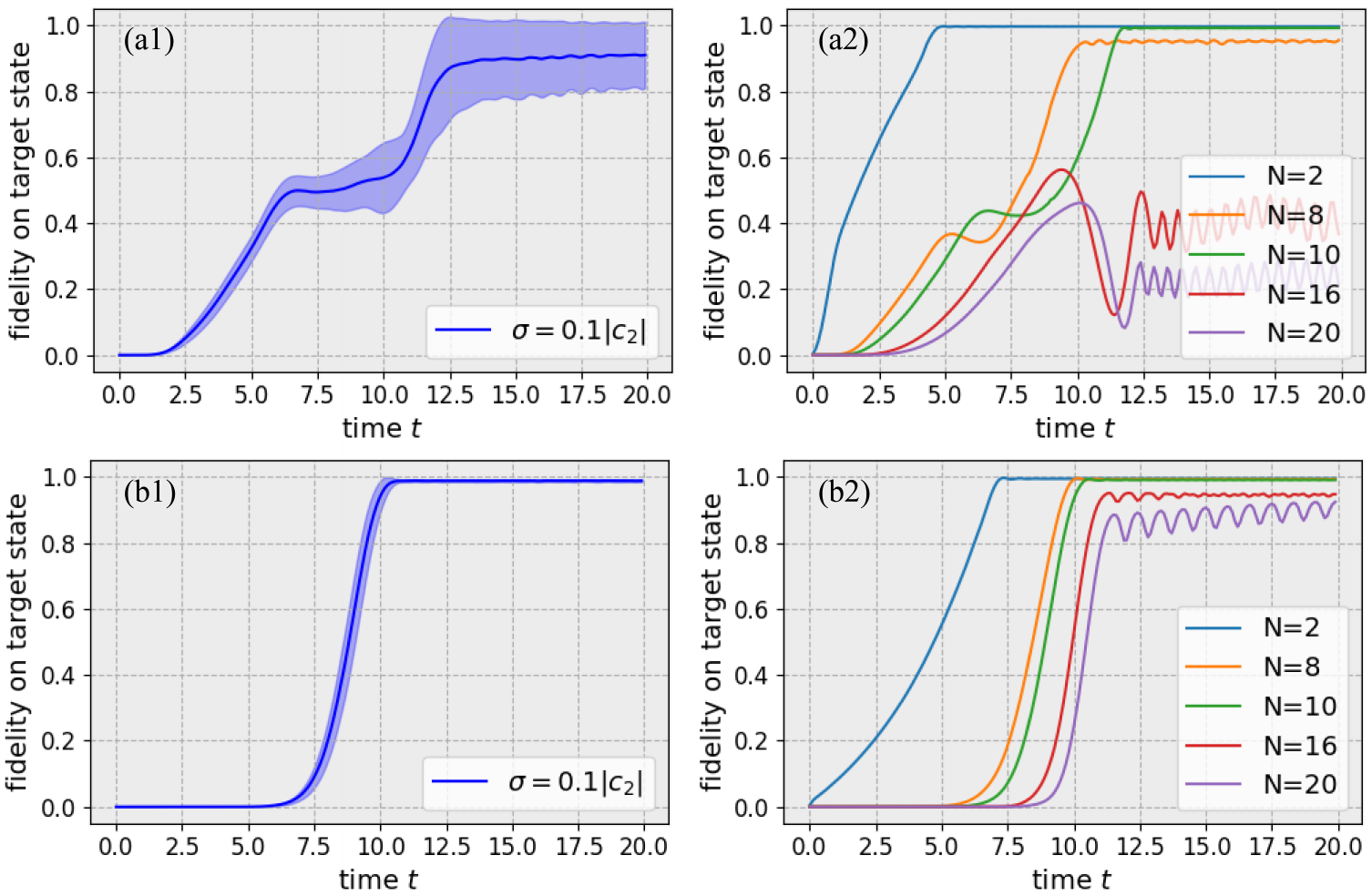}
\caption{(a1,b1)stability of policy $\pi_s$ and $\pi_g$ added by Gaussian white noise. The results are extracted from 100 sampled trajectories. The solid blue line is the average value of fidelity and the shaded band represents the standard deviation. (a2,b2)generalization capability of policy $\pi_s$ and $\pi_g$ learned from $N=10$  without noise.}
\label{sma_N10_stability}
\end{figure*}

	We further consider the stability and generalization capability of the polices learned by RL agent. We assume that the protocol $q(t)$ has a Gaussian type white noise
\begin{eqnarray}
q(t) = \frac{1}{\sqrt{2\pi}\sigma}e^{-\frac{(x-\mu(t))^2}{2\sigma^2}},
\label{sma_N10_noise}
\end{eqnarray}
where $\mu(t)$ is the exact value given by the policy and $\sigma$ indicates the strength of noise. Figs.(\ref{sma_N10_stability})(a1,b1) show the evolution of fidelity under noise (\ref{sma_N10_noise}) with $\sigma=0.1|c_2|$. The width of the shaded band represents the standard deviation of fidelity over 100 sampled trajectories. It is shown that $\pi_g$ is much more stable than $\pi_s$ even the protocol seems to be oscillating. We also apply the policy learned from $N=10$ to other particle number case to observe its generalization capability. As shown in Figs.(\ref{sma_N10_stability})(a2,b2), $\pi_g$ also has better performance than $\pi_s$. For $N<10$, both $\pi_s$ and $\pi_g$ are applicable because the dimension of Hilbert space is smaller than that of $N=10$. A policy learned in higher dimension should be available in its subspace. For $N>10$, only $\pi_g$ works because the state space is explored more thoroughly and true physical rules are learned under its setup. On the other hand, we notice that policy can be generalized to different particle number only when we use physical observables as state representation because the dimension of state input is fixed.

\section{Conclusion}\label{section-V}
	In conclusion, we use deep reinforcement learning to optimize twin-Fock state preparation in spin-1 atomic system. It not only gives better control protocols than some traditional methods but also presents a general scheme to convert a physical dynamical system into a standard RL task. We have noticed that state representation is important to the physical interpretability of final policy and reward signal should not be sparse. One central problem of RL algorithm is exploration and exploit. Only when the environment is effectively explored and experiences data are collected, agent could learn optimal policy without falling into local minimums (maximums). In our problem, we expect that using randomized or multiple initial state in each training episode could help to ease such problems and makes the polices being more robust and easily to be generalized to different environments.
	
	In section-\ref{section-IV} we also point out two major problems in many body system. For the first problem caused by POMDP, it could be partially solved by using recurrent RL \cite{2015arXiv150706527H} or using multiple steps trajectory as state input \cite{mnih2015human}. The second problem of sparse reward is even more essential and might be promoted by using hierachical reinforcement learning \cite{dietterich2000hierarchical,barto2003recent} framework.

\bibliography{RL_reference}

\begin{thebibliography}{35}%
\makeatletter
\providecommand \@ifxundefined [1]{%
 \@ifx{#1\undefined}
}%
\providecommand \@ifnum [1]{%
 \ifnum #1\expandafter \@firstoftwo
 \else \expandafter \@secondoftwo
 \fi
}%
\providecommand \@ifx [1]{%
 \ifx #1\expandafter \@firstoftwo
 \else \expandafter \@secondoftwo
 \fi
}%
\providecommand \natexlab [1]{#1}%
\providecommand \enquote  [1]{``#1''}%
\providecommand \bibnamefont  [1]{#1}%
\providecommand \bibfnamefont [1]{#1}%
\providecommand \citenamefont [1]{#1}%
\providecommand \href@noop [0]{\@secondoftwo}%
\providecommand \href [0]{\begingroup \@sanitize@url \@href}%
\providecommand \@href[1]{\@@startlink{#1}\@@href}%
\providecommand \@@href[1]{\endgroup#1\@@endlink}%
\providecommand \@sanitize@url [0]{\catcode `\\12\catcode `\$12\catcode
  `\&12\catcode `\#12\catcode `\^12\catcode `\_12\catcode `\%12\relax}%
\providecommand \@@startlink[1]{}%
\providecommand \@@endlink[0]{}%
\providecommand \url  [0]{\begingroup\@sanitize@url \@url }%
\providecommand \@url [1]{\endgroup\@href {#1}{\urlprefix }}%
\providecommand \urlprefix  [0]{URL }%
\providecommand \Eprint [0]{\href }%
\providecommand \doibase [0]{http://dx.doi.org/}%
\providecommand \selectlanguage [0]{\@gobble}%
\providecommand \bibinfo  [0]{\@secondoftwo}%
\providecommand \bibfield  [0]{\@secondoftwo}%
\providecommand \translation [1]{[#1]}%
\providecommand \BibitemOpen [0]{}%
\providecommand \bibitemStop [0]{}%
\providecommand \bibitemNoStop [0]{.\EOS\space}%
\providecommand \EOS [0]{\spacefactor3000\relax}%
\providecommand \BibitemShut  [1]{\csname bibitem#1\endcsname}%
\let\auto@bib@innerbib\@empty
\bibitem [{\citenamefont {Brouzos}\ \emph {et~al.}(2015)\citenamefont
  {Brouzos}, \citenamefont {Streltsov}, \citenamefont {Negretti}, \citenamefont
  {Said}, \citenamefont {Caneva}, \citenamefont {Montangero},\ and\
  \citenamefont {Calarco}}]{PhysRevA.92.062110}%
  \BibitemOpen
  \bibfield  {author} {\bibinfo {author} {\bibfnamefont {I.}~\bibnamefont
  {Brouzos}}, \bibinfo {author} {\bibfnamefont {A.~I.}\ \bibnamefont
  {Streltsov}}, \bibinfo {author} {\bibfnamefont {A.}~\bibnamefont {Negretti}},
  \bibinfo {author} {\bibfnamefont {R.~S.}\ \bibnamefont {Said}}, \bibinfo
  {author} {\bibfnamefont {T.}~\bibnamefont {Caneva}}, \bibinfo {author}
  {\bibfnamefont {S.}~\bibnamefont {Montangero}}, \ and\ \bibinfo {author}
  {\bibfnamefont {T.}~\bibnamefont {Calarco}},\ }\href {\doibase
  10.1103/PhysRevA.92.062110} {\bibfield  {journal} {\bibinfo  {journal} {Phys.
  Rev. A}\ }\textbf {\bibinfo {volume} {92}},\ \bibinfo {pages} {062110}
  (\bibinfo {year} {2015})}\BibitemShut {NoStop}%
\bibitem [{\citenamefont {Caruso}\ \emph {et~al.}(2012)\citenamefont {Caruso},
  \citenamefont {Montangero}, \citenamefont {Calarco}, \citenamefont {Huelga},\
  and\ \citenamefont {Plenio}}]{PhysRevA.85.042331}%
  \BibitemOpen
  \bibfield  {author} {\bibinfo {author} {\bibfnamefont {F.}~\bibnamefont
  {Caruso}}, \bibinfo {author} {\bibfnamefont {S.}~\bibnamefont {Montangero}},
  \bibinfo {author} {\bibfnamefont {T.}~\bibnamefont {Calarco}}, \bibinfo
  {author} {\bibfnamefont {S.~F.}\ \bibnamefont {Huelga}}, \ and\ \bibinfo
  {author} {\bibfnamefont {M.~B.}\ \bibnamefont {Plenio}},\ }\href {\doibase
  10.1103/PhysRevA.85.042331} {\bibfield  {journal} {\bibinfo  {journal} {Phys.
  Rev. A}\ }\textbf {\bibinfo {volume} {85}},\ \bibinfo {pages} {042331}
  (\bibinfo {year} {2012})}\BibitemShut {NoStop}%
\bibitem [{\citenamefont {Brif}\ \emph {et~al.}(2010)\citenamefont {Brif},
  \citenamefont {Chakrabarti},\ and\ \citenamefont {Rabitz}}]{Brif_2010}%
  \BibitemOpen
  \bibfield  {author} {\bibinfo {author} {\bibfnamefont {C.}~\bibnamefont
  {Brif}}, \bibinfo {author} {\bibfnamefont {R.}~\bibnamefont {Chakrabarti}}, \
  and\ \bibinfo {author} {\bibfnamefont {H.}~\bibnamefont {Rabitz}},\ }\href
  {\doibase 10.1088/1367-2630/12/7/075008} {\bibfield  {journal} {\bibinfo
  {journal} {New Journal of Physics}\ }\textbf {\bibinfo {volume} {12}},\
  \bibinfo {pages} {075008} (\bibinfo {year} {2010})}\BibitemShut {NoStop}%
\bibitem [{\citenamefont {Beltrani}\ \emph {et~al.}(2011)\citenamefont
  {Beltrani}, \citenamefont {Dominy}, \citenamefont {Ho},\ and\ \citenamefont
  {Rabitz}}]{beltrani2011exploring}%
  \BibitemOpen
  \bibfield  {author} {\bibinfo {author} {\bibfnamefont {V.}~\bibnamefont
  {Beltrani}}, \bibinfo {author} {\bibfnamefont {J.}~\bibnamefont {Dominy}},
  \bibinfo {author} {\bibfnamefont {T.-S.}\ \bibnamefont {Ho}}, \ and\ \bibinfo
  {author} {\bibfnamefont {H.}~\bibnamefont {Rabitz}},\ }\href@noop {}
  {\bibfield  {journal} {\bibinfo  {journal} {The Journal of chemical physics}\
  }\textbf {\bibinfo {volume} {134}},\ \bibinfo {pages} {194106} (\bibinfo
  {year} {2011})}\BibitemShut {NoStop}%
\bibitem [{\citenamefont {Yan}\ \emph {et~al.}(2014)\citenamefont {Yan},
  \citenamefont {Hocker}, \citenamefont {Long}, \citenamefont {Ho},\ and\
  \citenamefont {Rabitz}}]{PhysRevA.89.063408}%
  \BibitemOpen
  \bibfield  {author} {\bibinfo {author} {\bibfnamefont {J.}~\bibnamefont
  {Yan}}, \bibinfo {author} {\bibfnamefont {D.}~\bibnamefont {Hocker}},
  \bibinfo {author} {\bibfnamefont {R.}~\bibnamefont {Long}}, \bibinfo {author}
  {\bibfnamefont {T.-S.}\ \bibnamefont {Ho}}, \ and\ \bibinfo {author}
  {\bibfnamefont {H.}~\bibnamefont {Rabitz}},\ }\href {\doibase
  10.1103/PhysRevA.89.063408} {\bibfield  {journal} {\bibinfo  {journal} {Phys.
  Rev. A}\ }\textbf {\bibinfo {volume} {89}},\ \bibinfo {pages} {063408}
  (\bibinfo {year} {2014})}\BibitemShut {NoStop}%
\bibitem [{\citenamefont {Rothman}\ \emph {et~al.}(2006)\citenamefont
  {Rothman}, \citenamefont {Ho},\ and\ \citenamefont
  {Rabitz}}]{PhysRevA.73.053401}%
  \BibitemOpen
  \bibfield  {author} {\bibinfo {author} {\bibfnamefont {A.}~\bibnamefont
  {Rothman}}, \bibinfo {author} {\bibfnamefont {T.-S.}\ \bibnamefont {Ho}}, \
  and\ \bibinfo {author} {\bibfnamefont {H.}~\bibnamefont {Rabitz}},\ }\href
  {\doibase 10.1103/PhysRevA.73.053401} {\bibfield  {journal} {\bibinfo
  {journal} {Phys. Rev. A}\ }\textbf {\bibinfo {volume} {73}},\ \bibinfo
  {pages} {053401} (\bibinfo {year} {2006})}\BibitemShut {NoStop}%
\bibitem [{\citenamefont {Rabitz}\ \emph {et~al.}(2006)\citenamefont {Rabitz},
  \citenamefont {Ho}, \citenamefont {Hsieh}, \citenamefont {Kosut},\ and\
  \citenamefont {Demiralp}}]{PhysRevA.74.012721}%
  \BibitemOpen
  \bibfield  {author} {\bibinfo {author} {\bibfnamefont {H.}~\bibnamefont
  {Rabitz}}, \bibinfo {author} {\bibfnamefont {T.-S.}\ \bibnamefont {Ho}},
  \bibinfo {author} {\bibfnamefont {M.}~\bibnamefont {Hsieh}}, \bibinfo
  {author} {\bibfnamefont {R.}~\bibnamefont {Kosut}}, \ and\ \bibinfo {author}
  {\bibfnamefont {M.}~\bibnamefont {Demiralp}},\ }\href {\doibase
  10.1103/PhysRevA.74.012721} {\bibfield  {journal} {\bibinfo  {journal} {Phys.
  Rev. A}\ }\textbf {\bibinfo {volume} {74}},\ \bibinfo {pages} {012721}
  (\bibinfo {year} {2006})}\BibitemShut {NoStop}%
\bibitem [{\citenamefont {Sala}\ \emph {et~al.}(2016)\citenamefont {Sala},
  \citenamefont {N\'u\~nez}, \citenamefont {Martorell}, \citenamefont
  {De~Sarlo}, \citenamefont {Zibold}, \citenamefont {Gerbier}, \citenamefont
  {Polls},\ and\ \citenamefont {Juli\'a-D\'{\i}az}}]{PhysRevA.94.043623}%
  \BibitemOpen
  \bibfield  {author} {\bibinfo {author} {\bibfnamefont {A.}~\bibnamefont
  {Sala}}, \bibinfo {author} {\bibfnamefont {D.~L.}\ \bibnamefont {N\'u\~nez}},
  \bibinfo {author} {\bibfnamefont {J.}~\bibnamefont {Martorell}}, \bibinfo
  {author} {\bibfnamefont {L.}~\bibnamefont {De~Sarlo}}, \bibinfo {author}
  {\bibfnamefont {T.}~\bibnamefont {Zibold}}, \bibinfo {author} {\bibfnamefont
  {F.}~\bibnamefont {Gerbier}}, \bibinfo {author} {\bibfnamefont
  {A.}~\bibnamefont {Polls}}, \ and\ \bibinfo {author} {\bibfnamefont
  {B.}~\bibnamefont {Juli\'a-D\'{\i}az}},\ }\href {\doibase
  10.1103/PhysRevA.94.043623} {\bibfield  {journal} {\bibinfo  {journal} {Phys.
  Rev. A}\ }\textbf {\bibinfo {volume} {94}},\ \bibinfo {pages} {043623}
  (\bibinfo {year} {2016})}\BibitemShut {NoStop}%
\bibitem [{\citenamefont {Campbell}\ \emph {et~al.}(2015)\citenamefont
  {Campbell}, \citenamefont {De~Chiara}, \citenamefont {Paternostro},
  \citenamefont {Palma},\ and\ \citenamefont {Fazio}}]{PhysRevLett.114.177206}%
  \BibitemOpen
  \bibfield  {author} {\bibinfo {author} {\bibfnamefont {S.}~\bibnamefont
  {Campbell}}, \bibinfo {author} {\bibfnamefont {G.}~\bibnamefont {De~Chiara}},
  \bibinfo {author} {\bibfnamefont {M.}~\bibnamefont {Paternostro}}, \bibinfo
  {author} {\bibfnamefont {G.~M.}\ \bibnamefont {Palma}}, \ and\ \bibinfo
  {author} {\bibfnamefont {R.}~\bibnamefont {Fazio}},\ }\href {\doibase
  10.1103/PhysRevLett.114.177206} {\bibfield  {journal} {\bibinfo  {journal}
  {Phys. Rev. Lett.}\ }\textbf {\bibinfo {volume} {114}},\ \bibinfo {pages}
  {177206} (\bibinfo {year} {2015})}\BibitemShut {NoStop}%
\bibitem [{\citenamefont {Opatrn\'y}\ \emph {et~al.}(2016)\citenamefont
  {Opatrn\'y}, \citenamefont {Saberi}, \citenamefont {Brion},\ and\
  \citenamefont {M\o{}lmer}}]{PhysRevA.93.023815}%
  \BibitemOpen
  \bibfield  {author} {\bibinfo {author} {\bibfnamefont {T.~c.~v.}\
  \bibnamefont {Opatrn\'y}}, \bibinfo {author} {\bibfnamefont {H.}~\bibnamefont
  {Saberi}}, \bibinfo {author} {\bibfnamefont {E.}~\bibnamefont {Brion}}, \
  and\ \bibinfo {author} {\bibfnamefont {K.}~\bibnamefont {M\o{}lmer}},\ }\href
  {\doibase 10.1103/PhysRevA.93.023815} {\bibfield  {journal} {\bibinfo
  {journal} {Phys. Rev. A}\ }\textbf {\bibinfo {volume} {93}},\ \bibinfo
  {pages} {023815} (\bibinfo {year} {2016})}\BibitemShut {NoStop}%
\bibitem [{\citenamefont {Huang}\ \emph {et~al.}(2018)\citenamefont {Huang},
  \citenamefont {Kang}, \citenamefont {Chen}, \citenamefont {Shi},
  \citenamefont {Song},\ and\ \citenamefont {Xia}}]{PhysRevA.97.012333}%
  \BibitemOpen
  \bibfield  {author} {\bibinfo {author} {\bibfnamefont {B.-H.}\ \bibnamefont
  {Huang}}, \bibinfo {author} {\bibfnamefont {Y.-H.}\ \bibnamefont {Kang}},
  \bibinfo {author} {\bibfnamefont {Y.-H.}\ \bibnamefont {Chen}}, \bibinfo
  {author} {\bibfnamefont {Z.-C.}\ \bibnamefont {Shi}}, \bibinfo {author}
  {\bibfnamefont {J.}~\bibnamefont {Song}}, \ and\ \bibinfo {author}
  {\bibfnamefont {Y.}~\bibnamefont {Xia}},\ }\href {\doibase
  10.1103/PhysRevA.97.012333} {\bibfield  {journal} {\bibinfo  {journal} {Phys.
  Rev. A}\ }\textbf {\bibinfo {volume} {97}},\ \bibinfo {pages} {012333}
  (\bibinfo {year} {2018})}\BibitemShut {NoStop}%
\bibitem [{\citenamefont {Sels}\ and\ \citenamefont
  {Polkovnikov}(2017)}]{SelsE3909}%
  \BibitemOpen
  \bibfield  {author} {\bibinfo {author} {\bibfnamefont {D.}~\bibnamefont
  {Sels}}\ and\ \bibinfo {author} {\bibfnamefont {A.}~\bibnamefont
  {Polkovnikov}},\ }\href {\doibase 10.1073/pnas.1619826114} {\bibfield
  {journal} {\bibinfo  {journal} {Proceedings of the National Academy of
  Sciences}\ }\textbf {\bibinfo {volume} {114}},\ \bibinfo {pages} {E3909}
  (\bibinfo {year} {2017})}\BibitemShut {NoStop}%
\bibitem [{\citenamefont {Rabitz}\ \emph {et~al.}(2004)\citenamefont {Rabitz},
  \citenamefont {Hsieh},\ and\ \citenamefont {Rosenthal}}]{Rabitz1998}%
  \BibitemOpen
  \bibfield  {author} {\bibinfo {author} {\bibfnamefont {H.~A.}\ \bibnamefont
  {Rabitz}}, \bibinfo {author} {\bibfnamefont {M.~M.}\ \bibnamefont {Hsieh}}, \
  and\ \bibinfo {author} {\bibfnamefont {C.~M.}\ \bibnamefont {Rosenthal}},\
  }\href {\doibase 10.1126/science.1093649} {\bibfield  {journal} {\bibinfo
  {journal} {Science}\ }\textbf {\bibinfo {volume} {303}},\ \bibinfo {pages}
  {1998} (\bibinfo {year} {2004})}\BibitemShut {NoStop}%
\bibitem [{\citenamefont {Mnih}\ \emph {et~al.}(2015)\citenamefont {Mnih},
  \citenamefont {Kavukcuoglu}, \citenamefont {Silver}, \citenamefont {Rusu},
  \citenamefont {Veness}, \citenamefont {Bellemare}, \citenamefont {Graves},
  \citenamefont {Riedmiller}, \citenamefont {Fidjeland},\ and\ \citenamefont
  {Ostrovski}}]{mnih2015human}%
  \BibitemOpen
  \bibfield  {author} {\bibinfo {author} {\bibfnamefont {V.}~\bibnamefont
  {Mnih}}, \bibinfo {author} {\bibfnamefont {K.}~\bibnamefont {Kavukcuoglu}},
  \bibinfo {author} {\bibfnamefont {D.}~\bibnamefont {Silver}}, \bibinfo
  {author} {\bibfnamefont {A.~A.}\ \bibnamefont {Rusu}}, \bibinfo {author}
  {\bibfnamefont {J.}~\bibnamefont {Veness}}, \bibinfo {author} {\bibfnamefont
  {M.~G.}\ \bibnamefont {Bellemare}}, \bibinfo {author} {\bibfnamefont
  {A.}~\bibnamefont {Graves}}, \bibinfo {author} {\bibfnamefont
  {M.}~\bibnamefont {Riedmiller}}, \bibinfo {author} {\bibfnamefont {A.~K.}\
  \bibnamefont {Fidjeland}}, \ and\ \bibinfo {author} {\bibfnamefont
  {G.}~\bibnamefont {Ostrovski}},\ }\href
  {https://www.nature.com/articles/nature14236/} {\bibfield  {journal}
  {\bibinfo  {journal} {Nature}\ }\textbf {\bibinfo {volume} {518}},\ \bibinfo
  {pages} {529} (\bibinfo {year} {2015})}\BibitemShut {NoStop}%
\bibitem [{\citenamefont {Silver}\ \emph {et~al.}(2016)\citenamefont {Silver},
  \citenamefont {Huang}, \citenamefont {Maddison}, \citenamefont {Guez},
  \citenamefont {Sifre}, \citenamefont {Van Den~Driessche}, \citenamefont
  {Schrittwieser}, \citenamefont {Antonoglou}, \citenamefont {Panneershelvam},\
  and\ \citenamefont {Lanctot}}]{silver2016mastering}%
  \BibitemOpen
  \bibfield  {author} {\bibinfo {author} {\bibfnamefont {D.}~\bibnamefont
  {Silver}}, \bibinfo {author} {\bibfnamefont {A.}~\bibnamefont {Huang}},
  \bibinfo {author} {\bibfnamefont {C.~J.}\ \bibnamefont {Maddison}}, \bibinfo
  {author} {\bibfnamefont {A.}~\bibnamefont {Guez}}, \bibinfo {author}
  {\bibfnamefont {L.}~\bibnamefont {Sifre}}, \bibinfo {author} {\bibfnamefont
  {G.}~\bibnamefont {Van Den~Driessche}}, \bibinfo {author} {\bibfnamefont
  {J.}~\bibnamefont {Schrittwieser}}, \bibinfo {author} {\bibfnamefont
  {I.}~\bibnamefont {Antonoglou}}, \bibinfo {author} {\bibfnamefont
  {V.}~\bibnamefont {Panneershelvam}}, \ and\ \bibinfo {author} {\bibfnamefont
  {M.}~\bibnamefont {Lanctot}},\ }\href
  {https://www.nature.com/articles/nature16961} {\bibfield  {journal} {\bibinfo
   {journal} {Nature}\ }\textbf {\bibinfo {volume} {529}},\ \bibinfo {pages}
  {484} (\bibinfo {year} {2016})}\BibitemShut {NoStop}%
\bibitem [{\citenamefont {Silver}\ \emph {et~al.}(2017)\citenamefont {Silver},
  \citenamefont {Hubert}, \citenamefont {Schrittwieser}, \citenamefont
  {Antonoglou}, \citenamefont {Lai}, \citenamefont {Guez}, \citenamefont
  {Lanctot}, \citenamefont {Sifre}, \citenamefont {Kumaran},\ and\
  \citenamefont {Graepel}}]{silver2017mastering}%
  \BibitemOpen
  \bibfield  {author} {\bibinfo {author} {\bibfnamefont {D.}~\bibnamefont
  {Silver}}, \bibinfo {author} {\bibfnamefont {T.}~\bibnamefont {Hubert}},
  \bibinfo {author} {\bibfnamefont {J.}~\bibnamefont {Schrittwieser}}, \bibinfo
  {author} {\bibfnamefont {I.}~\bibnamefont {Antonoglou}}, \bibinfo {author}
  {\bibfnamefont {M.}~\bibnamefont {Lai}}, \bibinfo {author} {\bibfnamefont
  {A.}~\bibnamefont {Guez}}, \bibinfo {author} {\bibfnamefont {M.}~\bibnamefont
  {Lanctot}}, \bibinfo {author} {\bibfnamefont {L.}~\bibnamefont {Sifre}},
  \bibinfo {author} {\bibfnamefont {D.}~\bibnamefont {Kumaran}}, \ and\
  \bibinfo {author} {\bibfnamefont {T.}~\bibnamefont {Graepel}},\ }\href
  {https://arxiv.org/abs/1712.01815} {\bibfield  {journal} {\bibinfo  {journal}
  {arXiv:1712.01815}\ } (\bibinfo {year} {2017})}\BibitemShut {NoStop}%
\bibitem [{\citenamefont {Bukov}(2018)}]{bukov2018reinforcement}%
  \BibitemOpen
  \bibfield  {author} {\bibinfo {author} {\bibfnamefont {M.}~\bibnamefont
  {Bukov}},\ }\href
  {https://journals.aps.org/prb/abstract/10.1103/PhysRevB.98.224305} {\bibfield
   {journal} {\bibinfo  {journal} {Physical Review B}\ }\textbf {\bibinfo
  {volume} {98}},\ \bibinfo {pages} {224305} (\bibinfo {year}
  {2018})}\BibitemShut {NoStop}%
\bibitem [{\citenamefont {Yu}\ \emph {et~al.}(2018)\citenamefont {Yu},
  \citenamefont {Albarran-Arriagada}, \citenamefont {Retamal}, \citenamefont
  {Wang}, \citenamefont {Liu}, \citenamefont {Ke}, \citenamefont {Meng},
  \citenamefont {Li}, \citenamefont {Tang},\ and\ \citenamefont
  {Solano}}]{yu2018reconstruction}%
  \BibitemOpen
  \bibfield  {author} {\bibinfo {author} {\bibfnamefont {S.}~\bibnamefont
  {Yu}}, \bibinfo {author} {\bibfnamefont {F.}~\bibnamefont
  {Albarran-Arriagada}}, \bibinfo {author} {\bibfnamefont {J.}~\bibnamefont
  {Retamal}}, \bibinfo {author} {\bibfnamefont {Y.-T.}\ \bibnamefont {Wang}},
  \bibinfo {author} {\bibfnamefont {W.}~\bibnamefont {Liu}}, \bibinfo {author}
  {\bibfnamefont {Z.-J.}\ \bibnamefont {Ke}}, \bibinfo {author} {\bibfnamefont
  {Y.}~\bibnamefont {Meng}}, \bibinfo {author} {\bibfnamefont {Z.-P.}\
  \bibnamefont {Li}}, \bibinfo {author} {\bibfnamefont {J.-S.}\ \bibnamefont
  {Tang}}, \ and\ \bibinfo {author} {\bibfnamefont {E.}~\bibnamefont
  {Solano}},\ }\href {https://arxiv.org/abs/1808.09241} {\bibfield  {journal}
  {\bibinfo  {journal} {arXiv:1808.09241}\ } (\bibinfo {year}
  {2018})}\BibitemShut {NoStop}%
\bibitem [{\citenamefont {Andreasson}\ \emph {et~al.}(2018)\citenamefont
  {Andreasson}, \citenamefont {Johansson}, \citenamefont {Liljestrand},\ and\
  \citenamefont {Granath}}]{andreasson2018quantum}%
  \BibitemOpen
  \bibfield  {author} {\bibinfo {author} {\bibfnamefont {P.}~\bibnamefont
  {Andreasson}}, \bibinfo {author} {\bibfnamefont {J.}~\bibnamefont
  {Johansson}}, \bibinfo {author} {\bibfnamefont {S.}~\bibnamefont
  {Liljestrand}}, \ and\ \bibinfo {author} {\bibfnamefont {M.}~\bibnamefont
  {Granath}},\ }\href {https://arxiv.org/abs/1811.12338} {\bibfield  {journal}
  {\bibinfo  {journal} {arXiv:1811.12338}\ } (\bibinfo {year}
  {2018})}\BibitemShut {NoStop}%
\bibitem [{\citenamefont {Herbert}\ and\ \citenamefont
  {Sengupta}(2018)}]{herbert2018using}%
  \BibitemOpen
  \bibfield  {author} {\bibinfo {author} {\bibfnamefont {S.}~\bibnamefont
  {Herbert}}\ and\ \bibinfo {author} {\bibfnamefont {A.}~\bibnamefont
  {Sengupta}},\ }\href {https://arxiv.org/abs/1812.11619} {\bibfield  {journal}
  {\bibinfo  {journal} {arXiv:1812.11619}\ } (\bibinfo {year}
  {2018})}\BibitemShut {NoStop}%
\bibitem [{\citenamefont {Lin}\ \emph {et~al.}(2018)\citenamefont {Lin},
  \citenamefont {Lai},\ and\ \citenamefont {Li}}]{lin2018reinforcement}%
  \BibitemOpen
  \bibfield  {author} {\bibinfo {author} {\bibfnamefont {J.}~\bibnamefont
  {Lin}}, \bibinfo {author} {\bibfnamefont {Z.~Y.}\ \bibnamefont {Lai}}, \ and\
  \bibinfo {author} {\bibfnamefont {X.}~\bibnamefont {Li}},\ }\href
  {https://arxiv.org/abs/1812.10797} {\bibfield  {journal} {\bibinfo  {journal}
  {arXiv:1812.10797}\ } (\bibinfo {year} {2018})}\BibitemShut {NoStop}%
\bibitem [{\citenamefont {Nautrup}\ \emph {et~al.}(2018)\citenamefont
  {Nautrup}, \citenamefont {Delfosse}, \citenamefont {Dunjko}, \citenamefont
  {Briegel},\ and\ \citenamefont {Friis}}]{nautrup2018optimizing}%
  \BibitemOpen
  \bibfield  {author} {\bibinfo {author} {\bibfnamefont {H.~P.}\ \bibnamefont
  {Nautrup}}, \bibinfo {author} {\bibfnamefont {N.}~\bibnamefont {Delfosse}},
  \bibinfo {author} {\bibfnamefont {V.}~\bibnamefont {Dunjko}}, \bibinfo
  {author} {\bibfnamefont {H.~J.}\ \bibnamefont {Briegel}}, \ and\ \bibinfo
  {author} {\bibfnamefont {N.}~\bibnamefont {Friis}},\ }\href
  {https://arxiv.org/abs/1812.08451} {\bibfield  {journal} {\bibinfo  {journal}
  {arXiv:1812.08451}\ } (\bibinfo {year} {2018})}\BibitemShut {NoStop}%
\bibitem [{\citenamefont {Sweke}\ \emph {et~al.}(2018)\citenamefont {Sweke},
  \citenamefont {Kesselring}, \citenamefont {van Nieuwenburg},\ and\
  \citenamefont {Eisert}}]{sweke2018reinforcement}%
  \BibitemOpen
  \bibfield  {author} {\bibinfo {author} {\bibfnamefont {R.}~\bibnamefont
  {Sweke}}, \bibinfo {author} {\bibfnamefont {M.~S.}\ \bibnamefont
  {Kesselring}}, \bibinfo {author} {\bibfnamefont {E.~P.}\ \bibnamefont {van
  Nieuwenburg}}, \ and\ \bibinfo {author} {\bibfnamefont {J.}~\bibnamefont
  {Eisert}},\ }\href {https://arxiv.org/abs/1810.07207} {\bibfield  {journal}
  {\bibinfo  {journal} {arXiv:1810.07207}\ } (\bibinfo {year}
  {2018})}\BibitemShut {NoStop}%
\bibitem [{\citenamefont {Bukov}\ \emph {et~al.}(2017)\citenamefont {Bukov},
  \citenamefont {Day}, \citenamefont {Sels}, \citenamefont {Weinberg},
  \citenamefont {Polkovnikov},\ and\ \citenamefont {Mehta}}]{bukov2017machine}%
  \BibitemOpen
  \bibfield  {author} {\bibinfo {author} {\bibfnamefont {M.}~\bibnamefont
  {Bukov}}, \bibinfo {author} {\bibfnamefont {A.~G.}\ \bibnamefont {Day}},
  \bibinfo {author} {\bibfnamefont {D.}~\bibnamefont {Sels}}, \bibinfo {author}
  {\bibfnamefont {P.}~\bibnamefont {Weinberg}}, \bibinfo {author}
  {\bibfnamefont {A.}~\bibnamefont {Polkovnikov}}, \ and\ \bibinfo {author}
  {\bibfnamefont {P.}~\bibnamefont {Mehta}},\ }\href
  {https://arxiv.org/abs/1705.00565} {\bibfield  {journal} {\bibinfo  {journal}
  {arXiv:1705.00565}\ } (\bibinfo {year} {2017})}\BibitemShut {NoStop}%
\bibitem [{\citenamefont {Paparo}\ \emph {et~al.}(2014)\citenamefont {Paparo},
  \citenamefont {Dunjko}, \citenamefont {Makmal}, \citenamefont
  {Martin-Delgado},\ and\ \citenamefont {Briegel}}]{PhysRevX.4.031002}%
  \BibitemOpen
  \bibfield  {author} {\bibinfo {author} {\bibfnamefont {G.~D.}\ \bibnamefont
  {Paparo}}, \bibinfo {author} {\bibfnamefont {V.}~\bibnamefont {Dunjko}},
  \bibinfo {author} {\bibfnamefont {A.}~\bibnamefont {Makmal}}, \bibinfo
  {author} {\bibfnamefont {M.~A.}\ \bibnamefont {Martin-Delgado}}, \ and\
  \bibinfo {author} {\bibfnamefont {H.~J.}\ \bibnamefont {Briegel}},\ }\href
  {\doibase 10.1103/PhysRevX.4.031002} {\bibfield  {journal} {\bibinfo
  {journal} {Phys. Rev. X}\ }\textbf {\bibinfo {volume} {4}},\ \bibinfo {pages}
  {031002} (\bibinfo {year} {2014})}\BibitemShut {NoStop}%
\bibitem [{\citenamefont {Kitagawa}\ and\ \citenamefont
  {Ueda}(1993)}]{PhysRevA.47.5138}%
  \BibitemOpen
  \bibfield  {author} {\bibinfo {author} {\bibfnamefont {M.}~\bibnamefont
  {Kitagawa}}\ and\ \bibinfo {author} {\bibfnamefont {M.}~\bibnamefont
  {Ueda}},\ }\href {\doibase 10.1103/PhysRevA.47.5138} {\bibfield  {journal}
  {\bibinfo  {journal} {Phys. Rev. A}\ }\textbf {\bibinfo {volume} {47}},\
  \bibinfo {pages} {5138} (\bibinfo {year} {1993})}\BibitemShut {NoStop}%
\bibitem [{\citenamefont {M\"ustecapl\ifmmode \imath \else \i
  \fi{}o\ifmmode~\breve{g}\else \u{g}\fi{}lu}\ \emph
  {et~al.}(2002)\citenamefont {M\"ustecapl\ifmmode \imath \else \i
  \fi{}o\ifmmode~\breve{g}\else \u{g}\fi{}lu}, \citenamefont {Zhang},\ and\
  \citenamefont {You}}]{PhysRevA.66.033611}%
  \BibitemOpen
  \bibfield  {author} {\bibinfo {author} {\bibfnamefont {O.~E.}\ \bibnamefont
  {M\"ustecapl\ifmmode \imath \else \i \fi{}o\ifmmode~\breve{g}\else
  \u{g}\fi{}lu}}, \bibinfo {author} {\bibfnamefont {M.}~\bibnamefont {Zhang}},
  \ and\ \bibinfo {author} {\bibfnamefont {L.}~\bibnamefont {You}},\ }\href
  {\doibase 10.1103/PhysRevA.66.033611} {\bibfield  {journal} {\bibinfo
  {journal} {Phys. Rev. A}\ }\textbf {\bibinfo {volume} {66}},\ \bibinfo
  {pages} {033611} (\bibinfo {year} {2002})}\BibitemShut {NoStop}%
\bibitem [{\citenamefont {L{\"u}cke}\ \emph {et~al.}(2011)\citenamefont
  {L{\"u}cke}, \citenamefont {Scherer}, \citenamefont {Kruse}, \citenamefont
  {Pezz{\'e}}, \citenamefont {Deuretzbacher}, \citenamefont {Hyllus},
  \citenamefont {Topic}, \citenamefont {Peise}, \citenamefont {Ertmer},
  \citenamefont {Arlt}, \citenamefont {Santos}, \citenamefont {Smerzi},\ and\
  \citenamefont {Klempt}}]{Lucke773}%
  \BibitemOpen
  \bibfield  {author} {\bibinfo {author} {\bibfnamefont {B.}~\bibnamefont
  {L{\"u}cke}}, \bibinfo {author} {\bibfnamefont {M.}~\bibnamefont {Scherer}},
  \bibinfo {author} {\bibfnamefont {J.}~\bibnamefont {Kruse}}, \bibinfo
  {author} {\bibfnamefont {L.}~\bibnamefont {Pezz{\'e}}}, \bibinfo {author}
  {\bibfnamefont {F.}~\bibnamefont {Deuretzbacher}}, \bibinfo {author}
  {\bibfnamefont {P.}~\bibnamefont {Hyllus}}, \bibinfo {author} {\bibfnamefont
  {O.}~\bibnamefont {Topic}}, \bibinfo {author} {\bibfnamefont
  {J.}~\bibnamefont {Peise}}, \bibinfo {author} {\bibfnamefont
  {W.}~\bibnamefont {Ertmer}}, \bibinfo {author} {\bibfnamefont
  {J.}~\bibnamefont {Arlt}}, \bibinfo {author} {\bibfnamefont {L.}~\bibnamefont
  {Santos}}, \bibinfo {author} {\bibfnamefont {A.}~\bibnamefont {Smerzi}}, \
  and\ \bibinfo {author} {\bibfnamefont {C.}~\bibnamefont {Klempt}},\ }\href
  {\doibase 10.1126/science.1208798} {\bibfield  {journal} {\bibinfo  {journal}
  {Science}\ }\textbf {\bibinfo {volume} {334}},\ \bibinfo {pages} {773}
  (\bibinfo {year} {2011})}\BibitemShut {NoStop}%
\bibitem [{\citenamefont {Gross}\ \emph {et~al.}(2011)\citenamefont {Gross},
  \citenamefont {Strobel}, \citenamefont {Nicklas}, \citenamefont {Zibold},
  \citenamefont {Bar-Gill}, \citenamefont {Kurizki},\ and\ \citenamefont
  {Oberthaler}}]{gross2011atomic}%
  \BibitemOpen
  \bibfield  {author} {\bibinfo {author} {\bibfnamefont {C.}~\bibnamefont
  {Gross}}, \bibinfo {author} {\bibfnamefont {H.}~\bibnamefont {Strobel}},
  \bibinfo {author} {\bibfnamefont {E.}~\bibnamefont {Nicklas}}, \bibinfo
  {author} {\bibfnamefont {T.}~\bibnamefont {Zibold}}, \bibinfo {author}
  {\bibfnamefont {N.}~\bibnamefont {Bar-Gill}}, \bibinfo {author}
  {\bibfnamefont {G.}~\bibnamefont {Kurizki}}, \ and\ \bibinfo {author}
  {\bibfnamefont {M.}~\bibnamefont {Oberthaler}},\ }\href
  {https://www.nature.com/articles/nature10654} {\bibfield  {journal} {\bibinfo
   {journal} {Nature}\ }\textbf {\bibinfo {volume} {480}},\ \bibinfo {pages}
  {219} (\bibinfo {year} {2011})}\BibitemShut {NoStop}%
\bibitem [{\citenamefont {Bookjans}\ \emph {et~al.}(2011)\citenamefont
  {Bookjans}, \citenamefont {Hamley},\ and\ \citenamefont
  {Chapman}}]{PhysRevLett.107.210406}%
  \BibitemOpen
  \bibfield  {author} {\bibinfo {author} {\bibfnamefont {E.~M.}\ \bibnamefont
  {Bookjans}}, \bibinfo {author} {\bibfnamefont {C.~D.}\ \bibnamefont
  {Hamley}}, \ and\ \bibinfo {author} {\bibfnamefont {M.~S.}\ \bibnamefont
  {Chapman}},\ }\href {\doibase 10.1103/PhysRevLett.107.210406} {\bibfield
  {journal} {\bibinfo  {journal} {Phys. Rev. Lett.}\ }\textbf {\bibinfo
  {volume} {107}},\ \bibinfo {pages} {210406} (\bibinfo {year}
  {2011})}\BibitemShut {NoStop}%
\bibitem [{\citenamefont {Luo}\ \emph {et~al.}(2017)\citenamefont {Luo},
  \citenamefont {Zou}, \citenamefont {Wu}, \citenamefont {Liu}, \citenamefont
  {Han}, \citenamefont {Tey},\ and\ \citenamefont {You}}]{Luo620}%
  \BibitemOpen
  \bibfield  {author} {\bibinfo {author} {\bibfnamefont {X.-Y.}\ \bibnamefont
  {Luo}}, \bibinfo {author} {\bibfnamefont {Y.-Q.}\ \bibnamefont {Zou}},
  \bibinfo {author} {\bibfnamefont {L.-N.}\ \bibnamefont {Wu}}, \bibinfo
  {author} {\bibfnamefont {Q.}~\bibnamefont {Liu}}, \bibinfo {author}
  {\bibfnamefont {M.-F.}\ \bibnamefont {Han}}, \bibinfo {author} {\bibfnamefont
  {M.~K.}\ \bibnamefont {Tey}}, \ and\ \bibinfo {author} {\bibfnamefont
  {L.}~\bibnamefont {You}},\ }\href {\doibase 10.1126/science.aag1106}
  {\bibfield  {journal} {\bibinfo  {journal} {Science}\ }\textbf {\bibinfo
  {volume} {355}},\ \bibinfo {pages} {620} (\bibinfo {year}
  {2017})}\BibitemShut {NoStop}%
\bibitem [{\citenamefont {Schulman}\ \emph {et~al.}(2017)\citenamefont
  {Schulman}, \citenamefont {Wolski}, \citenamefont {Dhariwal}, \citenamefont
  {Radford},\ and\ \citenamefont {Klimov}}]{schulman2017proximal}%
  \BibitemOpen
  \bibfield  {author} {\bibinfo {author} {\bibfnamefont {J.}~\bibnamefont
  {Schulman}}, \bibinfo {author} {\bibfnamefont {F.}~\bibnamefont {Wolski}},
  \bibinfo {author} {\bibfnamefont {P.}~\bibnamefont {Dhariwal}}, \bibinfo
  {author} {\bibfnamefont {A.}~\bibnamefont {Radford}}, \ and\ \bibinfo
  {author} {\bibfnamefont {O.}~\bibnamefont {Klimov}},\ }\href
  {https://arxiv.org/abs/1707.06347} {\bibfield  {journal} {\bibinfo  {journal}
  {arXiv:1707.06347}\ } (\bibinfo {year} {2017})}\BibitemShut {NoStop}%
\bibitem [{\citenamefont {Hausknecht}\ and\ \citenamefont
  {Stone}(2015)}]{2015arXiv150706527H}%
  \BibitemOpen
  \bibfield  {author} {\bibinfo {author} {\bibfnamefont {M.}~\bibnamefont
  {Hausknecht}}\ and\ \bibinfo {author} {\bibfnamefont {P.}~\bibnamefont
  {Stone}},\ }\href {https://arxiv.org/abs/1507.06527} {\bibfield  {journal}
  {\bibinfo  {journal} {arXiv:1507.06527}\ } (\bibinfo {year}
  {2015})}\BibitemShut {NoStop}%
\bibitem [{\citenamefont {Dietterich}(2000)}]{dietterich2000hierarchical}%
  \BibitemOpen
  \bibfield  {author} {\bibinfo {author} {\bibfnamefont {T.~G.}\ \bibnamefont
  {Dietterich}},\ }\href
  {https://www.jair.org/index.php/jair/article/view/10266} {\bibfield
  {journal} {\bibinfo  {journal} {Journal of Artificial Intelligence Research}\
  }\textbf {\bibinfo {volume} {13}},\ \bibinfo {pages} {227} (\bibinfo {year}
  {2000})}\BibitemShut {NoStop}%
\bibitem [{\citenamefont {Barto}\ and\ \citenamefont
  {Mahadevan}(2003)}]{barto2003recent}%
  \BibitemOpen
  \bibfield  {author} {\bibinfo {author} {\bibfnamefont {A.~G.}\ \bibnamefont
  {Barto}}\ and\ \bibinfo {author} {\bibfnamefont {S.}~\bibnamefont
  {Mahadevan}},\ }\href
  {https://link.springer.com/article/10.1023/A:1022140919877} {\bibfield
  {journal} {\bibinfo  {journal} {Discrete event dynamic systems}\ }\textbf
  {\bibinfo {volume} {13}},\ \bibinfo {pages} {41} (\bibinfo {year}
  {2003})}\BibitemShut {NoStop}%
\end{thebibliography}%

\end{document}